# Nucleon-Nucleon Interaction in the Singlet State at Low Energy.

S.B. Borzakov*

Frank Laboratory of Neutron Physics, JINR

Abstract

The nucleon-nucleon interaction at low energy usually is described with help of effective range theory. The scattering parameters are scattering length and effective range. The scattering length for triplet state is positive and S-matrix pole corresponds to the deuteron. The scattering length for singlet *np*-interaction is negative and S-matrix pole corresponds to the virtual state, but physical sense of the virtual level is unclear. It is possible to describe the scattering and the radiative capture with help of a Breit-Wigner resonance with negative energy.

So the conclusion about nonexistence of bound singlet state is not correct probably.

A number of theoretical works which based on quantum chromodynamics show that singlet metastable state of the (*n+p*) system exists. The mass of this state is approximately equal to sum of masses for two nucleons. Their predictions support the representations about negative resonance.

We treated singlet deuteron as dibaryon resonance with the mass which is a little lower than sum of neutron and proton masses, but bigger than deuteron mass. We propose the direct experiments to search for bound state of the *np* – system with the zero momentum (singlet deuteron).

*serborz@list.ru

# Nucleon-Nucleon interaction in $^1S_0$ state at low energy.

I. Nucleon-Nucleon interaction at very low energy.
   1. The scattering amplitude, effective range theory and the S-matrix poles.
   2. The NN interaction at low energy as Breit-Wigner resonance.
   3. Microscopic calculations. Dibaryon resonances.

II. The possible experiments to search for $^1S_0$ dibaryon state.

### I. The nucleon-nucleon interaction at very low energy.

Total momentum for two nucleons is the sum of spins and orbital momentum: $\vec{J} = \vec{S}_1 + \vec{S}_2 + \vec{L}$. Nucleon-nucleon interaction with orbital momentum **L= 0** is possible in two states: $J^+ = 1$ (triplet state), or $J^- = 0$ (singlet state). According to the Pauli principle the interaction between two identical nucleons is possible in the singlet state only.

The total scattering cross-section is equal to the sum of the scattering cross-sections for triplet ($\sigma_t$) and singlet states ($\sigma_s$) multiplied by statistical weights:

$$\sigma = \frac{2 \cdot J^+ + 1}{2 \cdot (2 \cdot I + 1)} \sigma_+ + \frac{2 \cdot J^- + 1}{2 \cdot (2 \cdot I + 1)} \sigma_- = \frac{3}{4} \sigma_t + \frac{1}{4} \sigma_s \qquad (1)$$

Scattering cross-section is proportional to square of the scattering amplitude: $\sigma_{scat} = 4 \cdot \pi \cdot |F|^2$

$$\sigma = \frac{3}{4} \sigma_t + \frac{1}{4} \sigma_s = \pi \cdot \left(3 \cdot |F_t|^2 + |F_s|^2\right) \qquad (2)$$

The scattering amplitude is well known from the scattering theory:

$$F = \frac{1}{k \cdot ctg\,\delta - i \cdot k} = \frac{1}{g(k) - i \cdot k} \qquad (3)$$

$\delta(k)$ - is the scattering phase.

Where **k** is the neutron momentum: $k = \frac{1}{\hbar}\sqrt{2 \cdot \mu \cdot E}$, $\mu = \frac{m_n \cdot m_p}{m_n + m_p}$ is reduced mass.

One can present the function **g(k)** as:

$$g(k) = -\frac{1}{a} + \frac{1}{2}\rho k^2 + Pk^4 + Qk^6 + ... \qquad (4)$$

The main scattering parameters are scattering length *a* and effective range *ρ*. The values of the scattering length and effective range for each spin state have been determined from number of experiments and are shown in the Table 1.

S-matrix is connected with scattering amplitude:

$$S = 1 + 2ikF(k) \qquad (5)$$

In the first approximation the effective range theory (ERT) is used for scattering amplitude:

$$F = \frac{1}{-\frac{1}{a} + \frac{1}{2}\rho k^2 - ik}, \qquad (6)$$

where *a* is scattering length and *ρ* is effective range.

If we know the values of scattering lengths and effective ranges one can determine the poles of the S-matrix for each spin state for negative energy. If $E_n < 0$ then $k = i \cdot \kappa$, where $\kappa$ is real value. When the Coulomb interaction is absent it necessary to solve the quadratic equation to search for poles:

$$-\frac{1}{a} - \frac{1}{2}\rho\kappa^2 + \kappa = 0 \qquad (7)$$

The two roots exist for each spin state:

$$\kappa_{1,2} = \frac{1}{\rho} \pm \sqrt{\frac{1}{\rho^2} - \frac{2}{a\rho}} \qquad (8)$$

Table 1. The experimental values of the scattering parameters.

|  | Scattering length, Fm | Effective range, Fm | P | Q |
|---|---|---|---|---|
| *pp* | -7.822 ± 0.004 | 2.83 ± 0.017 | 0.051 ± 0.014 | 0.028 ± 0.013 |
| *np* ($^1S_0$) | -23.719 ± 0.013 | 2.76 ± 0.05 | - | - |
| *nn* | -18.7 ± 0.6 -16 | - | - | - |
| *np* ($^3S_1$) | 5.414 ± 0.005 | 1.750 ± 0.005 | 0.13 ± 0.09 | - |

The results of the calculations are shown in the Table 2.

Table 2.

| State | $\kappa_{1,2}$, 1/Fm | E, MeV | Comment |
|---|---|---|---|
| $^3S_1$ ($a_t > 0$) | 0.232 | - 2.225 | Deuteron |
|  | 0.911 | - 34.4 | Physical sense is unclear |
| $^1S_0$ ($a_s < 0$) | - 0.044 | - 0.080 | Virtual level |
|  | 1.20 | - 59.6 | Physical sense is unclear |

The S-matrix poles for triplet state are situated at imaginary axis in upper part of the plane. This situation describes bound states. The corresponding binding energy for one from the poles is equal to measured deuteron binding energy 2224 keV. The scattering length for $^1S_0$-state is negative ($a_s$ = -23.72 Fm) and one of the poles is situated in low part of $k$-plane. This situation corresponds to so named virtual (or antibound) level. The physical sense of a virtual level is unclear. His wave function increase infinitely at big distance. But the state energy is negative.

From the idea of the presence of the virtual level, it was concluded that the *np* system in the singlet state is not bound. This result is interpreted in such a way that the interaction potential is not deep enough for the formation of a bound state [1]. This conclusion was included in a number of monographs and textbooks. However, it follows from the analysis described above that the ERT is not a strict theory and, therefore, such a conclusion cannot be recognized as proved and needs experimental verification.

But the number of questions arise concerning this model:

    a.    How to determine the energy of the virtual level experimentally?

    b.    It is well known that sign of real part of the scattering amplitude can't be determined from experiment. V. de Alfaro, T. Regge wrote: "Only if there is no bound state capable to account for the low-energy cross-section one is entitled to give definite statements about the existence of antibound states" (from "Potential scattering", North-Holland Publishing Company- Amsterdam, 1965, p.72).

    c.    This model does not include the radiative capture.

d. The description of the interaction of the neutrons with another lightest nuclei (for example, deuterons and $^3$He) is possible with help of ERT also. But there is no an accordance between the S-matrix poles and bound states.

Thus, the statement that there is an unambiguous relationship between the poles of the S-matrix and the bound states is not experimentally confirmed. Therefore, the idea that there is no bound state of two nucleons with spin $0^+$ needs experimental verification.

**2. The NN interaction at low energy as Breit-Wigner resonance. R-matrix theory.**

S.T. Ma in 1953 y. proposed another description of the *NN* scattering – with help of *R*-matrix theory [3] The similar model is discussed in [4].

It is well known that not only the elastic scattering but radiative capture is possible from singlet *np* state. It is possible to describe the capture with help of complex scattering length: $A_s = a_s - i \cdot b$ [5].

It is possible to rewrite the expression for scattering amplitude:

$$F = \frac{1}{-\frac{1}{a-ib} + \frac{1}{2}\rho k^2 - ik} = \frac{1}{2k} \frac{\frac{4}{\rho}k}{k^2 - \frac{2}{\rho a} - i\frac{1}{2}(\frac{4k}{\rho} + \frac{4b}{\rho a^2})}$$

$$= \frac{1}{2k} \frac{\Gamma_n}{E - E_r - i\frac{1}{2}(\Gamma_n + \Gamma_\gamma)}$$

Neutron width increases proportional to square root of energy: $\Gamma_n = \frac{4k}{\rho} \propto \sqrt{E_n}$, according to *R*-matrix theory. Resonance energy is equal to: $E_r = \frac{2}{ar}\frac{\hbar^2}{2\mu}$ in accordance with the work by S.T. Ma.

We add the radiative width $\Gamma_\gamma = \frac{4b}{\rho a^2} \cdot \frac{\hbar^2}{2\mu} = 41.47 MeV \cdot Fm^2$

The obtained formula equivalent to Breit-Wigner formula, which usually describes the resonance.

It is easy to show that the radiative capture cross-section is describes by the formula $\sigma_{n\gamma} = \frac{\pi}{k}b$ and follows to the low $\frac{1}{v} = \frac{1}{\sqrt{E_n}}$ at $E_n \to 0$ (v is the neutron velocity). The radiative cross-section for thermal neutrons is known with very good accuracy $\sigma_{n,\gamma}(0.025eV) = (334 \pm 0.5)$ mb. One can calculate the value of the parameter $b = 2.7 \cdot 10^{-4}$ Fm.

Because of the singlet scattering length is negative the resonance energy is negative also. Thus the idea of existence of metastable state with the mass which is little less than sum of the masses for neutron and proton. Using the known values of the scattering length and effective radius, we obtain the following estimates:  $E_r$ = -1,3 МэВ; $\Gamma_n$ (1 эВ) = 10 кэВ; $\Gamma_\gamma$ = 20 эВ.

It is better to introduce the excitation energy: $E^* = B_d + E_{n,cm}$. When $E^* < B_d$ ($B_d$ - deuteron binding energy) the different variants of the model are possible. The first – neutron width is equal to zero. The second – the value of the neutron width is imaginary. In this case the S-matrix pole corresponds to the virtual level. But virtual level has radiative width in this case. It means that it is real quasistationary level.

This model allows us to describe the scattering, radiative capture, photodisintegration and scattering of gamma-quanta by deuteron with help of three parameters: resonance energy, neutron width and radiative width by unit way.

This approach is widely used in neutron physics to describe the interaction of neutrons with various nuclei. This approach has led to success in describing the interaction of neutrons with $^3$He using the $^4$He level lying below the threshold of the decay of $^4$He into $^3$He and neutron [6]. Subsequently, this level was discovered in a direct experiment - the scattering of protons by tritium.

### 3. Microscopic calculations.

The phenomenological description of the nucleon-nucleon interaction at low energies was discussed above. Of course, the question arises about a microscopic description of the properties of the NN system. It is now generally accepted that nucleons are composed of quarks. Immediately after the creation of the quark theory, calculations of states with different quantum numbers for two nucleons appeared. In the work of F. Dyson and N. Kh. Huong, it was found that the masses of the triplet and singlet states are approximately equal [7].

K. Maltman and N. Izgur calculated the energies of the triplet and states for two nucleons in the frame of the six quark model. The result for singlet state is $E(^1S_0)$ = (-0.4 ± 0.4) MeV, for base triplet state (deuteron) $B_d$ = -2.9 MeV [8]. It is possible to consider this work as prediction of the deuteron exited state. Here the energy is counted from the sum of masses for two nucleons.

A.N. Ivanov et al. predict the metastable singlet deuteron with binding energy $B(^1S_0) = 79 \pm 12$ keV. The calculations have been made on the base of Nambu-Iona-Lasinio model [9].

A similar idea is developed by R. Hackenburg from the Brookhaven Laboratory [10]. He developed a method for calculating the processes of elastic scattering, radiation capture and scattering of gamma quanta by deuterons, based on the diagram technique taking into account the intermediate dibaryon state. According to R. Hackenburg, there is a quasistationary singlet state with a lifetime of $4 \cdot 10^{-17}$ sec and with a mass of 66 keV less than the sum of the masses of the neutron and proton.

Calculations of Yamazaki et al. in the quenched lattice quantum chromodynamics lead to the conclusion that not only the triplet but singlet deuteron state should be bound [11].

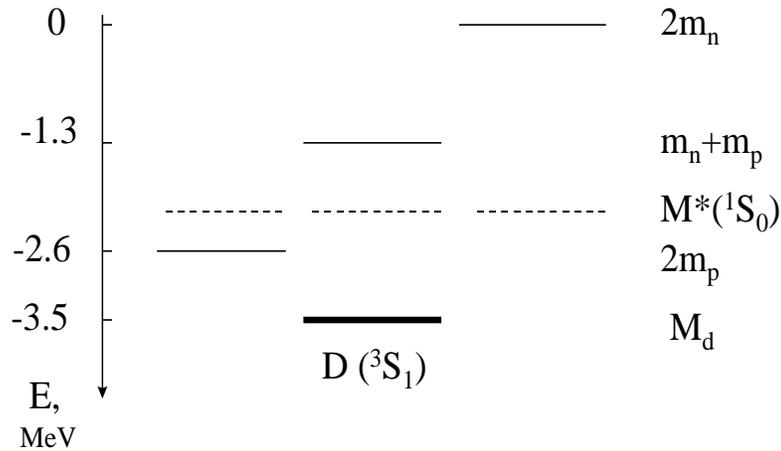

Fig. 1 Comparison of the masses of nucleon pairs.

So it is possible to imagine the existence of the dibaryon with isotope spin T = 1 and spin is equal to zero. The mass of this dibaryon is less than sum of neutron's and proton masses ($m_n+m_p$), but bigger than deuteron mass. The difference between sums of masses for different pairs of nucleons is shown on the Fig. 1. It is well known that the neutron mass bigger than proton mass on 1.293 MeV. The hypothetical dibaryon is indicated by a dashed line. It could manifest itself in the interaction of neutrons with protons, neutrons with neutrons and protons with protons, as well as in the interaction of nucleons with antinucleons.

Thus, the question of the existence of a bound singlet state of two nucleons is of exceptional interest. Closely connected with it is the problem of the existence of a dineutron, that is, a nucleus consisting of two neutrons. According to the principle of isotopic invariance, the absence of a bound level in a deuteron leads to the absence of a dineutron in nature. However, as follows from the above, the statement that the deuteron does not have a singlet bound state does not rely on any strict theoretical model. Moreover, modern theoretical models predict the bound singlet states for two nucleons. In my opinion the existence of the dineutron does not contradict to any fundamental physical laws.

II. The possible experiments to search for $^1S_0$ dibaryon state.

The question of the existence of a dineutron was first raised by N. Feather in 1948 [12]. He estimated the possible lifetime of a dineutron with respect to beta decay of 1-5 seconds. Dozens of papers are devoted to the searches for a dineutron now.

According to N. Feather maximal binding energy of the dineutron can be obtained from the condition the dineutron decay to deuteron, electron and antineutrino.

The masses of the nucleon-nucleon pairs (***pp, np, nn***) are shown at the Fig 1. It is known that the neutron mass is bigger than proton mass on 1.293 MeV. Neutron and proton образуют the связанное состояние - дейтрон с энергией связи $B_d$ = 2.224 МэВ, total momentum $J^\pi = 1^+$, isotopic spin T = 0. From these data and using the condition

$$m(^2n)=2m_n-B_2 > m_d+m_e, = m_n+m_p-B_d+m_e, \qquad (2)$$

(where $m_d$ is the deuteron mass, $m_e$ is the electron mass and $B_2$ is the dineutron binding energy), one can obtained the maximal binding energy of the dineutron:

$$B_2 = 2m_n - m(^2n) <= m_n-m_p+B_d-m_e = 3.01 \text{ MeV} \qquad (3)$$

An activation technique was often used to search for neutral nuclei. Its essence lies in the fact that the induced activity is studied, which appears when a neutral particle is captured by some stable isotope. In the experiment, a sufficiently thick filter is placed between the source of neutral particles and the detector, which absorbs all charged particles but passes neutral ones. The first indication of the existence of a dineutron was obtained in the work of M. Sakisata and M. Tomita, who studied the reaction. They used the activation method. As an indicator, $^{27}$Al was used. During capture of a dineutron, $^{29}$Al activity should be observed. The tritium-titanium target was bombarded by deuterons with the energy of 160-185 keV. After irradiation, induced activity with half-lives of 6.6 minutes and 18 minutes was detected. The authors concluded that activity from 6.6 min. caused by the $^{29}$Al decay, which appeared as a result of the capture of a dineutron, and the binding energy E ($^2$n) = 2.90 - 3.01 MeV was estimated [13].

C. Detraz [14] measured the activity, which had arisen after irradiation of $^{235}$U by protons with the energy 24 GeV, and concluded that there is emission of neutral nuclei. The detector of neutral nuclei was the reaction of capture of two neutrons: $^2n+^{70}Zn \rightarrow ^{72}Zn$.

The clearest observation of the dineutron and singlet deuteron have been obtained by Bochkarev et al. [15]. They study the decay of the exited states (2$^+$) of the nuclei $^6$He, $^6$Li and $^6$Be. The amplitude spectra of alpha-particles emitted from these states have been measured. The very narrow peaks were observed in cases of $^6$He and $^6$Li decays. These facts indicate to the decays into two particles: alpha-particle and singlet deuteron in the first case and alpha-particle and dineutron in the second case. It is impossible to describe these peaks as two nucleon interaction in the final state.

Set and Parker studied the interaction of the mesons with nuclei $^6$Li and $^9$Be [16]. They measured the spectra of mesons and protons arising in the next reactions:

a) $^6$Li ($\pi^-$, $\pi^+$) $^6$H;  b) $^9$Be ($\pi^-$, p) $^8$He;  c) $^6$Li ($\pi^-$, p) $^5$H.

The measurements have been made at the energies of incident mesons 220 MeV (for reaction ***a***) and 125 MeV (reactions ***b*** and ***c***). The unstable nuclei were the final products of all three reactions. The yield of the registered particles, depending on the effective mass, is better consistent with experimental data, assuming the formation of a dineutron. The dineutron binding energy was assumed to equal to zero.

Recently, experiments with radioactive beams have been of great interest. The angular distribution of alpha particles in the scattering of the $^6$He on $^4$He have been measured [17]. The results show that the process with transfer of the two neutron pair has very big probability. The authors conclude the bound dineutron exists into the $^6$He. The existing of the dineutron can explain this fact also.

In our time there are many investigations with beams of unstable nuclei. F.M. Marques et al. observed 6 events of the appearance of tetraneutrons in the reaction with $^{14}$Be nuclei [18].

A. Spyrou et al. observed the dineutron emission in the decay of $^{16}$Be [19]. Analogous process has been observed in the reaction $^{26}$O – $^{24}$O +2n [20].

I.M. Kadenko et al. proposed new method to search for dineutron [21].

Search for the singlet deuteron.

Existence of the bound state singlet deuteron could be evidenced by two-step gamma-ray transition $^3S_1$ (continuum) → $^1S_0$ (metastable) → $^3S_1$ (ground state) in addition to the direct one

$^1S_0 \to {}^3S_1$ with energy 2223.25 keV. The experiment was performed at the prompt gamma activation analysis facility of the Budapest Neutron Center. The beam of cold neutrons was extracted from the cold source of the reactor. The neutron flux density was $10^8$ n/(cm$^2$·sec). The polyethylene target was located between two gamma-ray detectors: a BGO shielded Compton suppressed coaxial HPGe detector and low energy HPGe detector. The peak area of the main transition 2223.25 keV was $2.8 \cdot 10^8$. The main difficulty in the interpretation of the experimental spectra was caused by the background of gamma-rays generated in the detector and material surrounding the target and the detector. Numerous peaks from neutron radiative capture by isotopes of Ge, Cl, Fe, Al, C, N etc. have been observed in measured spectra.

The conclusion has been made that there is no evidence for two-photon transition with one of gamma-rays in the region 2100-2210 keV. The branching ratio relative to main transition is R < $6 \cdot 10^{-6}$ or cross-section < 2 µb (two standard deviations) [22].

The peak with the energy 2212.9 keV have been observed in these spectra. The branching ratio relative to main transition R = $2 \cdot 10^{-4}$. The authors of this work have decided that this peak can be explained the following process: when photo effect occurs at detector surface the Ge atoms emit roentgen quanta which fly from the sensitive volume. As a result the part of energy (approximately 10 keV) loosed. This effect is known for detectors which have small volume and for low energy of gamma quanta. But for the detector used in this work the probability of this effect is very small. We have carried out the measurements using similar HPGe detector and $^{24}$Na as radioactive source to search for this effect. $^{24}$Na emit two gamma lines with energies 1368 and 2754 keV. We didn't observe the peaks with energies on 10 keV lower (1358 and 2744 keV) at the level of $10^{-5}$ as compare with peak areas of main transitions.

So it is possible to consider the peak 2212.9 keV as an indication of the singlet deuteron existence.

New experiments are needed to search for the dineutron and singlet deuteron. It is necessary to register the coincidence of two gamma rays. The sensitivity to the process under study can be increased many times by suppressing the contribution from Compton scattered gamma quanta.

Literature.

1. L.D. Landau, E. M. Lifshitz, "Quantum mechanics: Non-Relativistic Theory", Pergamon Press, 1977.
2. V. de Alfaro, T. Regge, "Potential scattering", North-Holland Publ. Comp., Amsterdam, 1965.